\documentclass[twocolumn,3p,times,procedia]{elsarticle}

\usepackage{ecrc}
\usepackage{graphicx}
\usepackage{titlesec}

\volume{00}

\firstpage{1}

\runauth{}

\jid{procs}

\CopyrightLine{2020}{Published by Elsevier Ltd.}

\usepackage{amssymb}
\usepackage{amsmath}
\usepackage{multicol}

\usepackage[figuresright]{rotating}
\usepackage{bm}
\usepackage{caption}

\usepackage{fancyhdr}

\fancyheadoffset[RO,EL]{0pt}
\usepackage{geometry}

\usepackage{multirow}
\usepackage{threeparttable}

\usepackage{amsmath}  

\usepackage[ruled]{algorithm2e}

\begin{document}

\begin{frontmatter}




\dochead{}
\title{
\begin{flushleft}
{\bf \Huge Comparison and Improvement for Delay Analysis Approaches: Theoretical  Models and Experimental Tests} 
\end{flushleft}
}
 %

\author[]{\bf \Large \leftline {Yue Hong Gao$^*$, Xiao Hong, Hao Tian Yang, Lu Chen, Xiao Nan Zhang}}

\address{\bf  \leftline {Wireless Theory and Wireless Technology Laboratory, Beijing University of Post and Telecommunication, Beijing 100876, China}

}

\cortext[]{Corresponding author. \emph{E-mail address}: yhgao@bupt.edu.cn (Y. Gao).}
\begin{abstract}

Computer network tends to be subjected to the proliferation of mobile demands and increasingly multifarious, therefore it poses a great challenge to guarantee the quality of network service. By designing the model according to different requirements, we may get some related indicators such as delay and packet loss rate in order to evaluate the quality of network service and verify the user data surface and capacity of the network environment. 
In this paper, we describe an analytical model based on the measurement for the delay of each packet passing through the single existing routers in the network environment. In previous studies, the emulation of real network service behaviors was generally under ideal condition. In our work, the test environment is built to get the relevant test results of the actual network, and the corresponding theoretical results are obtained by our model. The test results are compared with the theoretical results, analyzed and corrected, in order to verify the feasibility of our analysis model for the performance analysis of the actual network. With this concern, calculation results are modified with different schemes to realize more precise calculation of delay boundary with the comparison with the experimental test results. The results show the analysis methods after the amendment can realistically estimate the performance of network element.

\end{abstract}


\end{frontmatter}


\section{Introduction}
The infrastructure and service flow of packet networks are becoming more and more complex [1,2,3]. Essentially, the quality of service is determined by the packet transmission service provided by the network. User satisfaction with the service can also reflect the quality of service. With the rapid development of computer network technology, quality of service has become an important feature of the Internet. The components used in these networks apply different scheduling methods and QoS specifications, such as traffic shaping, pre initialization and so on. For more complex packet forwarding performance, many advanced network evaluation methods are based on mathematical models, depending on service performance indicators such as delay, packet loss rate to evaluate the actual performance of network nodes. In the modern network environment, the calculation and analysis of data surface performance in different stages of the network life cycle has become the focus of many scholars. However, due to the internal structure, scheduling and buffer attributes of the switch, it is very complex to consider the details of hardware to describe the network node model. Therefore, the mathematical model to simulate the actual equipment is critical. To analyze and interpret the performance of service devices and improve their models, Literature [4] proposes a methodology to identify the factors contributing to single-hop delay. In particular, the largest part of single-hop delay experienced by a packet is not due to queueing, but rather to the processing unit and transmission of the packet across the switch fabric. 

The performance of various hardware [4] and (virtualized) software [5] routers is also a subject for many scholars. For the sake of the analysis of traffic scheduling algorithms in  packet networks, Literature [6] presents a general Qos model, called Latency-Rate servers, involving two parameters the latency and the allocated rate. Thus, the theory of Latency-Rate servers solves the problem of calculating upper bound on end-to-end delay and buffer requirements in a network of servers. The Latency-Rate model also constrains the parameter values under different schedules and the applicability of this model is verified. 

In Literature [7], the delay results show that the method performance derived by new service curve considering the fact that transmission occurs at full line rate is better than the conventional algorithm with a known transmission rate [9] bottomed on ideal assumptions, especially in the part that the packet length is diverse. The information on the transmission rate is exploited to provide a bound on the delay at a FIFO system that improves on the foregone algorithm. 

Modeling realtime systems with shared resources is a challenge due to the transition to multi-processor system. Literature [8] compares two means to conduct simulation for servers in the actual network environment and details the analysis that models the  processing unit to performance bound and latency-rate (LR) servers using temporal abstractions. 

The application of network calculus has made the modeling and analysis techniques developed versatile and uncertain and applicable to network and cloud implementations and heterogeneous networks and computing systems coexisting in the cloud environment. Reference [9] proposed a new modeling method for composite network computing service capability and a comprehensive network computing service performance evaluation technology based on network calculus.

The characterization of data services in networks is also a key poser technologically for many scholars. It is a challenge to accurately describe business data streams using mathematical models. [10] models the data flow arrival process and service process under different distributions individually, which proposed a method to analyze the access delay bounds in Internet utilizing stochastic network calculus. In addition to the tremendous changes in data service, hardware chips are also increasingly difficult to meet demand, so bandwidth is gradually becoming the bottleneck of the Internet. Due to the development of hardware chip is sophisticated, long cycle, high cost and the decline of network service quality will lead to the unavailability of data service and affect the user experience, how to build service model is the focus of research.

The main content of this paper is to evaluate the quality of network service more practically. The analysis methods of data model and
service model in network calculus are applied to single node scenario to realize the more precise calculation of delay boundary through the comparison with the test results in this paper. The structure of this paper is organized as follows: In section II, we review the related
previous studies and present the traditional general system model. In section  \uppercase\expandafter{\romannumeral3} we present forwarding process of the single node model in detail and propose a algorithm on basis of measurement to derive server performance parameters for improved service curve. Moreover, our first interest will be on modeling servers, and then on modeling flows of data. Then, we propose modification schemes for the arrival process affected by the various factors in real network environment in section \uppercase\expandafter{\romannumeral4} and corrections are made to some of the problems that occurred during the testing process. Thus, we will provide parameter calibration methods which in specific to the rate and burst constrain of the incoming traffic flow. Section IV summarizes the research work.

\section{A Common System Model}
In this paper, we apply network calculus and evaluate the performance of routers and network entities by means of experimental measurements. Network calculus is a theory to analyze the boundary of traffic delay and the boundary of server data backlog in computer and communication networks. The transform from the complex nonlinear network system to the analyzable linear system by using alternate algebra is the fundamental of network calculus [11]. The mathematical modeling is a abstraction of the real network system to obtain the indexes of performance to evaluate the network performance. Diversified and complex businesses are abstracted as arrival curves, service processes are abstracted as service curves, and network quality can be assessed using the relevant Qos boundaries in network calculus. 

Then we will introduce the arrival curve and service curve [12] in traditional deterministic network calculus to lay the foundation for its subsequent theoretical analysis as in Figure 1. The packet at the head of the queue is transmitted at a constant rate $C$ and the integrated system offer a service curve $\beta$ to the incoming flow. All packets of the incoming flow is packetized and then we assume that the flow is constrained by an arrival curve $\alpha$.

Consider a server with arrival process $A(t)$ and departure process $D(t)$ offers a (deterministic) service curve $\beta$ to the input flow A. Suppose A has a (deterministic) arrival curve $\alpha$. Then, the delay $d$ of the flow A is bounded by
$$
d(A,D)\leqslant h(\alpha,\beta)
$$
which $h(\alpha,\beta)$ represents the maximum horizontal distance between them. 

\begin{figure}[htbp]
\centerline
{
\includegraphics[height=2.8cm,width=7.6cm]
{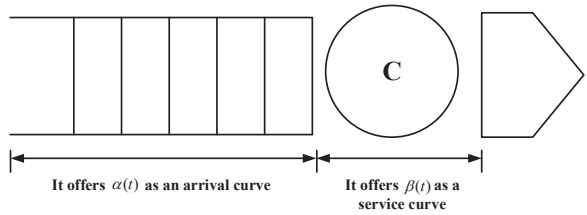}
}
\caption{Common system model}
\label{fig1}
\end{figure}

Generally, the delay $d$ (only considered queue waiting delay) has been derived by service curve $\beta_{R,e}:t\mapsto R[t-e]^{+}$ with a nominal rate $C$ and no additional processing delay and arrival curve $\alpha$ without consideration of port speed limit in the ideal system [5]:
$$
d=d_{\{R=C,e=0\}}=\frac{Packet Length}{C}
$$

In most cases, however, the transmittion system is at the actual rate $R$ lower than nominal rate $C$ with the existence of extra dealy on account of the occupancy to the system service rate for other preprocessor subsystems. The behavior of data flow that crosses the realistic network element is hard to crystallized. Thus, the modification schemes of arrival curve and service curve are proposed in the next section.  

\section{Improved Method for the Service Curve Estimation}
Assuming the case where a single server is fed with one flow in the ideal system, the delay can be derived by service curve with a nominal rate $C$. However, the parameters for arrival curve and service curve, such as arrival rate and service rate, are often influenced by external factors in actual test environment.

\begin{figure}[htbp]
\centerline
{
\includegraphics[height=2.6cm,width=7.6cm]
{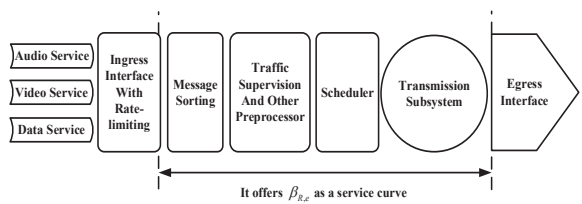}
}
\caption{Qos processing technology flow}
\label{fig2}
\end{figure}

The processing technology of QoS including diffluence of messages, administration of congestion, avoidance of congestion, wardship and plastic of fluxis detailed in the Figure 2. A data flow composed of multiple data flows enters the server from the input port. After the process of token bucket speed limit, de-multiplexing, classification and tagging, the income flow is scheduled to wait for service in different queue caches according to its own priority, and then flows to downstream devices after token bucket shaping. Token bucket speed limit essentially plays a role of traffic regulation, restricting traffic to a specific bandwidth. If the traffic exceeds the bandwidth limit, the traffic beyond the rated bandwidth will be discarded, so as to avoid an unrestricted bandwidth occupation by a user's business. Queue scheduling refers to determining the forwarding order of messages according to the corresponding scheduling algorithm when the network is congested, in the interest of ensure that the network can resume normal and stable state at the fastest speed. Traffic reshaping actively adjusts the output rate of streams so that they can be more smoothly transmitted to downstream devices to avoid unnecessary packet dropping and traffic congestion. 

The performance of the server in the actual network will also be affected by factors such as packet length, load size, and so on. In many cases, the rate $R$ is much less than nominal rate $C$ specially when the transmission capacity $C$ is shared between this scheduler system and other preprocessor subsystems.

\subsection{Parameters Estimation Based On Measurement}

To estimate the service curve when the server is actually working, a method to depict server behaviors is proposed in this section. Considering various factors such as switch internal structure, scheduling and buffer attributes, we simplify the server into a model agreed upon by service rate $R$ which represents the throughput per unit time and error term $e$ defined as additional device delays caused by processing unit, such as static delays for forwarding (flow instruction delays, static delays for host network cards), and dynamic delays (queue delays, delays for protocol stacks). In order to ensure the accuracy of subsequent network analysis, it is necessary to depict a more accurate service curve $\beta_{R,e}:t\mapsto R[t-e]^{+}$ so as to better assess network performance [6] as illustrated in Figure 3.

This section describes the measurement methods used in this paper, the parameters for service curve can be estimated by investigating the arriving and departing packets within backlog period. 

\begin{figure}[htbp]
\centerline
{
\includegraphics[height=5.4cm,width=7cm]
{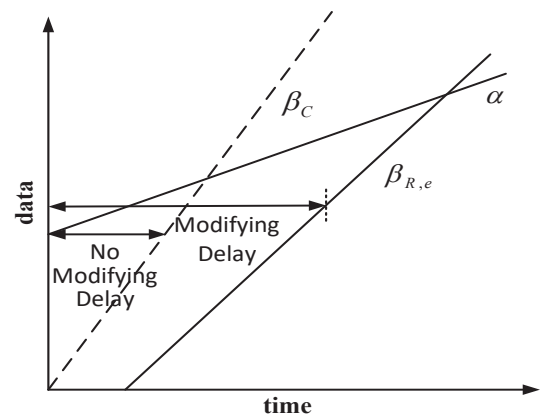}
}
\caption{Delay bound under modified model}
\label{fig2}
\end{figure}

The principle of service model is as follows: after a burst of packets enter the switch and cause congestion, called backlog period $B(t)$, such that the arrival process $A(t)$ and departure process $D(t)$ holds [10] 
$$B(t)=A(t)-D(t)\geqslant 0$$

The ideal server set to serve at a nominal rate $C$ is fed with the flow with fixed packet length $L$. Among the packets with arrival time $t_i$ and departure time $t_i^*$, the theoretical departure time of the first patcket is
$$t_1^*=t_1+\frac{L}{R}$$
\par\setlength\parindent{1em}The theoretical departure time of the second packet depending on the arrival time of the second packet and the departure time of the first packet from the server, is given by
$$
t_2^*=max\{t_2,t_1^*\}+\frac{L}{R}
$$
\par\setlength\parindent{1em}Thus, the the virtual finishing time of packet $i$ can be derived by
$$
t_i^*=max\{t_i,t_{i-1}^*\}+\frac{L}{R}
$$

Due to the actual service rate is often not up to the nominal rate, the actual departure time of packet $T_i^*$ is often later than theoretical departure time $t_i^*$. To simulate the server more accurately, we need to get the actual rate and error factor for service curve which satisfies
\begin{equation}
T_i^*\leqslant t_i^*+e
\end{equation}
where $T_i^*$ denotes the departure time and $t_i^*$ the virtual finishing time. With the increase of item $i$, the actual service rate $R$ value decreases from the theoretical rate $C$, and $T_i^*$ will exceeds $t_i^*$. On condition that the value of $T_i^*-t_i^*$ keep increasing, it is necessary to continue decreasing the $R$ value until error term that satisfies equation (1) for any packet becomes stable. Algorithm 1 shows the procedure how we get the parameters for modifying service curve $\beta_{R,e}:t\mapsto R[t-e]^{+}$.

\renewcommand{\arraystretch}{1.2}
\begin{table}[htbp]
\begin{threeparttable}
\begin{tabular}{|c|c|c|c|c|}
\hline
\multicolumn{2}{|c|}{Packet   Length} & 256B & 512B & 1500B \\ \hline
\multirow{4}{*}{\begin{tabular}[c]{@{}c@{}}IO delay (us) \\ under different\\ server  loads\\ (*1Gbps)\end{tabular}} &
  20\% &
  \multirow{4}{*}{2.4} &
  \multirow{4}{*}{2.4} &
  2.4 \\ \cline{2-2} \cline{5-5} 
                & 50\%                &         &         & 2.4      \\ \cline{2-2} \cline{5-5} 
                & 80\%                &         &         & 3.6      \\ \cline{2-2} \cline{5-5} 
                & 100\%               &         &         & 3.6      \\ \hline
\multicolumn{2}{|c|}{Maximum}         & 2.4     & 2.4     & 3.6      \\ \hline
\end{tabular}
 \begin{tablenotes}
        \footnotesize
        \item * The maximum transmission capacity of the forwarding-packet instrument is approximately 96\%.
      \end{tablenotes}
  \end{threeparttable}
\caption*{Tab. 1: IO delay in testbed environment}
\end{table}

Theoretically, the correction of service rate $R$ and error term $e$ has been completed. However, there is a larger gap between the measurement data and the theoretical calculated value. Thus, through reverting the entire test process, we found that the delay factor calculated in the above algorithm does not include the IO delay (i.e. the delay of the packet from applications to disk, which is almost fixed). The maximum of IO delay for the specified packet length under different loads is added to the error term $e$ in the algorithm in which case both the packet length and the load will affect the IO delay as shown in Table 1.

\begin{algorithm}
\caption{Parameter Estimation based on measurement}  
  \KwIn{the actual arriving time and depaturing time of each packet $T_1,T_2,...,T_n$ and $T_1^*,T_2^*,...,T_n^*$ through measurement }  
  \KwOut{System rate $R$ and error term $e$ for service curve $\beta_{R,e}$ after corrective scheme}  
  $R \gets \textit{nominal rate $C$}$\;
  \textbf{loop}:\;
  decrease $R$\;
  $t_1^* \gets \textit{$t_1+\frac{L}{R}$}$\;
  $ e_1 \gets \textit{$T_1^*-t_1^*$}$\;
  \For{$i = 2;i \le n;i++$}  
  {  
    $t_i^* \gets \textit{$max\{t_i^*,t_{i-1}^*\}+\frac{L}{R}$}$\; 
      \If{$T_i^*-t_i^* \ge T_{i-1}^*-t_{i-1}^*$}  
      {  
        \textbf{goto} loop\;  
      }
      \Else  
      {
       $ e_{i} \gets \textit{$T_{i}^*-t_{i}^*$}$\;
      }
  }
$e$ $\gets e_i$ among $\{e_1,e_2,e_3...e_n \}$ satisfies $T_i^*\leqslant t_i^*+e$ for all $T_i$ and $t_i$\; 
\textbf{return} System rate $R$ and error term $e$\;  
\end{algorithm}

The impact of variable packet length on the performance of improved model change as in Table 2, where the error term includes IO delay. In general, the guaranteed rate $R$ for service curve varies directly with packet length and error term $e$ is not directly related to the change of packet length. Thus, it is verified that the performance of service devices in the real network environment is  affected by the packet length. 

\begin{table}[htbp]
\begin{tabular}{|c|c|c|}
\hline
\begin{tabular}[c]{@{}c@{}}Packet \\ Length\end{tabular} &
  \begin{tabular}[c]{@{}c@{}}Service Rate(Gbps)\end{tabular} &
  \begin{tabular}[c]{@{}c@{}}Error Term(us)\end{tabular} \\ \hline
256B  & 885.95 & 4.20 \\ \hline
512B  & 918.19 & 4.20 \\ \hline
1500B & 941.21 & 5.00   \\ \hline
\end{tabular}
\caption*{Tab. 2: Parameters for service curve under different package length}
\end{table}

\subsection{Analysis of the Single-Hop Delay}

To illustrate this and keep things simple, consider a periodic flow of fixed packet length $l$ with rate $r$ and burst $b$ that crosses a server offering a service cuerve $\beta$ with rate $R$ and error term $e$. Assuming a server with arrival process $A(t)$ and departure process $D(t)$ offers a (deterministic) service curve $\beta$ to the input flow A. Suppose A has a (deterministic) arrival curve $\alpha_{r,b} :t\mapsto rt+b$. Then, the delay $d$ of the flow A is bounded by [11]
\begin{equation}
d(A,D) \leqslant h(\alpha_{r,b},\beta_{R,e}) = \frac{b}{R}+e
\end{equation}

The end-to-end delay is a kind of indexes for assessment of quality of network service. This metric can also be measured by the start time and deadline for a message. Total completion time of packet $i$ can be given by:
$$
T(i)=T_{queue}(i)+T_{proc}(i)+T_{trans}(i)
$$
where $T(i)$ represents the total completion time, $T_{wait}(i)$ indicates the queueing time as in equation (2), $T_{proc}(i)$ is the packet processing time like error term $e$ and $T_{trans}(i)$ means the transmission time. 

\begin{figure}[htbp]
\centerline
{
\includegraphics[height=5.8cm,width=6.8cm]
{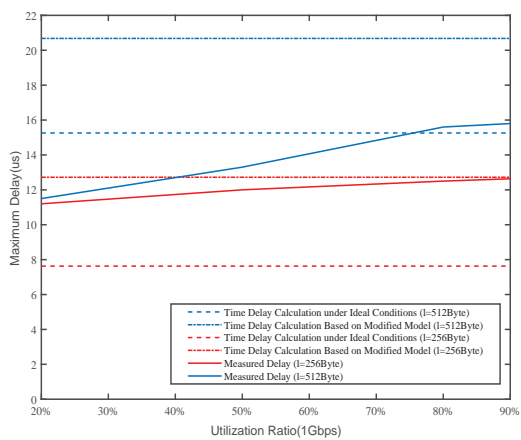}
}
\caption{Maximum delay with different utilization in the case where one single server is fed with one flow (burst = 3 packet length)}
\label{fig5}
\end{figure}

Our experiments were set up by using a simulation tool MATLAB to realize performance analysis for our end-to-end delay model. Figure 4 shows the maximum delay under different utilization with diverse packet length. As seen in Figure 4, the delay boundary calculated in the ideal case do not meet the test values under all loads. 

\renewcommand{\arraystretch}{1.2}
\begin{table}[htbp]
\begin{threeparttable}
\begin{tabular}{|c|c|c|c|c|c|}

\hline
\multirow{2}{*}{\begin{tabular}[c]{@{}c@{}}Parameter \\ configuration\end{tabular}} &
  \multicolumn{4}{c|}{\begin{tabular}[c]{@{}c@{}}Packet   burst\\ (*Packet length)\end{tabular}} &
  \multirow{2}{*}{$\sigma^2$} \\ \cline{2-5}
       & 1    & 3    & 5    & 8    &       \\ \hline
[256B, 50\%] & 12.0 & 12.9 & 11.8 & 11.6 & 0.3 \\ \hline
[512B, 50\%] & 13.3 & 13.9 & 14.0 & 12.8 & 0.3 \\ \hline
[512B, 80\%] & 15.4 & 15.4 & 15.2 & 15.4 & 0.1 \\ \hline
[1500B, 20\%] & 18.9 & 21.5 & 19.7 & 21.5 & 1.7 \\ \hline
\end{tabular}
 \begin{tablenotes}
        \footnotesize
        \item 1 Test parameter configuration is referred to as [Packet length, Server load].
				\item 2 $\sigma^2$ represents the variance of the maximum delay.
      \end{tablenotes}
  \end{threeparttable}
\caption*{Tab. 3: Maximum delay under test environment with various parameter configuration}

\end{table}

To tackle this problem, the model put forward in this section effectively modifies the delay boundaries so that all measured delay results are within the calculated delay range. For instance, we assume the input flow packet length is 256 Byte for single server model. The emulation results show that the delay is longer than 7.6us calculated under ideal condition, and all measured values change within the boundary computed based on modified model. Similarly, by comparing the delay of packets crossing through the switch with different packet length and the measured delay, it can be found that the calculated value is closer to the measured value when the packet length is shorter. Thus, the fact that the performance of server in real network environment is lower than the nominal capacity is verified. However, by observing the measured data in Figure 4, we find that the load intensity of the data stream has some effect on the delay, and that the consequence of burst packet length does not cause the delay to increase stepwise as illustrated in Table 3.

The data flow serverd by the server is regarded as a burst packet flow with ideal periodic change in this section. Nevertheless, in the actual test environment, the income flow to the server cannot arrive strictly according to the rule of periodic behavior. To improve model sensitivity to server load intensity and burst, further modifications in our model analysis will be actualized based on the arrival curve in the next section, so as to make the model closer to the actual network behavior.

\section{Method For Arrival Curve Under Real Scenario}
The test delay results in the last section illustrate that the current modified model does not respond to the change of server load, and the linearly increase of burst traffic does not make the test delay increase in accordance with a certain rule. So far, we  conjecture that these phenomena can be attributed to the link speed mechanism of the switch port or the fact that periodic burst packets do not meet the burst condition when they are generated from the source. 

In practice, the modle will change to some extent under constraint condition of port link-speed. With this concern, we put forward a correction scheme to deal with this porblem. Thus, it is critical to get the modified arrival curve $\alpha$. In this section, we will compare the delay results derived by the three arrival curve of the peroidic flow with the consideration of port speed limit and the burst condition for flow from the source as prerequisite.

\subsection{Pervasive Arrival Curve for Periodic Flow}
In previous research, we found that arrival curves have many various mathematical representations. Considering the complexity and diversity of data service, a universal arrvial curve $\alpha_{r,b} :t\mapsto rt+b$ is applied to our model. 

\par Considering the constraint of link-speed, a 4-tuple $(p,l,r,b)$ is used to specify traffic, where $p$ denotes
the link speed, $l$ the arriving flow packet length, $r$ the arriving flow rate, and $b$ the arriving flow burst. 

\begin{figure}[htbp]
\centerline
{
\includegraphics[height=5.2cm,width=7cm]
{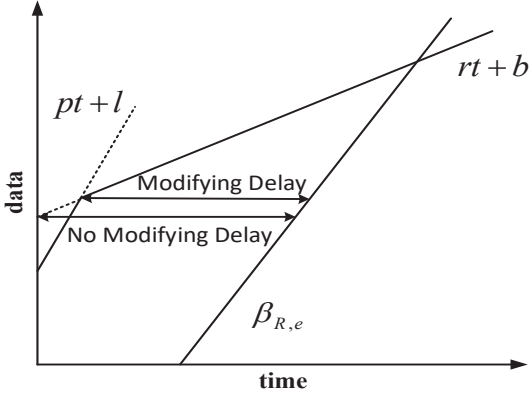}
}
\caption{Modifying delay vs. no modifying delay of arrival curves under link-speed constraint with 4-tuple model}
\label{fig3}
\end{figure}

With such a traffic specification, the flow has an arrival curve as illustrated in Figure 5:
\begin{equation}
\begin{aligned}
\alpha_{p,l,r,b}(t)&=\alpha_{p\land r,l\land b}(t)\\&=(pt+l)\land(rt+b)\\&=min\{pt+l,rt+b\}
\end{aligned}
\end{equation}

Comparing with the delay that would be obtained without considering the link-speed effect, we can derive the improved delay bound of arrving flow by equation (2) and (3) as follow:
\begin{equation}
\begin{aligned}
d(A,D)&\leqslant h(\alpha_{p,l,r,b},\beta_{R,e})\\&=h((pt+l)\land(rt+b),R[t-e]^{+})\\&=\frac{b}{R}+e-\frac{(R-r)(b-l)}{(p-r)R}
\end{aligned}
\end{equation}
which has to be compared to the delay bound without link-speed modifying:
$$
h(\alpha_{r,b},\beta_{R,e})=\frac{b}{R}+e
$$

The factor $\frac{(R-r)(b-l)}{(p-r)R}$ can be interpreted as a reduction factor that the effect of link-speed on the delay boundary becomes quantifiable. This reduction promotes the maximum delay boundary closer to actual delay.

\subsection{Strict Arrival Curve for Periodic Flow}
Since the server is fed with the periodic flow, a more strict arrival curve is adopted to achieve a higher accuracy for dealy boundary. In addition to depicting periodic flow as $\alpha_{p,l,r,b}$ with rate $r$ and burst $b$, more accurate arrival curve $\alpha_{p,l,T,b}$ can be obtained as in Figure 6. Suppose that a flow with arrival process $A(t)$ offers a arrival curve $\alpha$, by isotony and associativity
of the convolution, for all $n>0$, 
$$
A\leqslant A\otimes\alpha^{n}
\Rightarrow \ \ A\leqslant inf_{n\in \mathbb{N}}A\otimes\alpha^{n}=A\otimes\alpha^{*}
$$
where $\alpha^{*}\overset{def}{=} \land_{i=1}^{+\infty}f^{i}$ is also a arrival curve of the flow [13]. 

Asumming a periodic flow $\kappa_{b,T}=b\lceil \frac{t}{T} \rceil$ of period $T$ with burst packet size $b$ that crosses a server with constant rate $R$ and error term $e$, a tighter arrival curve in light of port speed can be computed as:
$$
\begin{aligned}
\alpha_{T,b,p,l}=\alpha_{T,b,p,l}^{*}&=((pt+l)\land b\lceil \frac{t}{T} \rceil)^{*}\\&=(pt+l)^{*}\otimes (b\lceil \frac{t}{T}\rceil)^{*}\\&=(pt+l)\otimes (b\lceil \frac{t}{T}\rceil)
\end{aligned}
$$

Thus, the delay boundary can be derived as:

\begin{equation}
\begin{aligned}
d(A,D)&\leqslant h(\alpha_{T,b},\beta_{R,e})\\&=\frac{b}{R}+e-\frac{b-l}{p}
\end{aligned}
\end{equation}

\begin{figure}[htbp]
\centerline
{
\includegraphics[height=4.6cm,width=7.0cm]
{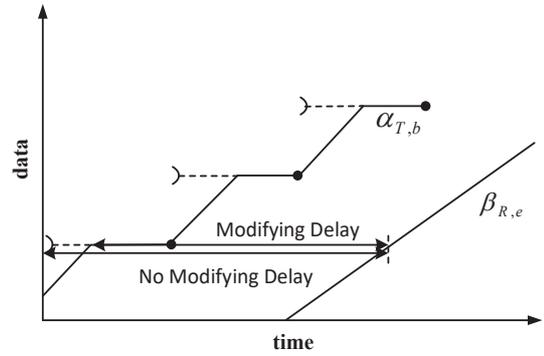}
}
\caption{Modifying vs. no modifying of arrival curves}
\label{fig7}
\end{figure}

In general, the universal model can improve the sensitivity of the calculated results to load changes, while a more rigorous arrival curve is more accurate in theory.

\subsection{Strict Arrival Curve for Data Flow in Actual Network Environment}
In addition to the effect of link speed, whether the generation of data flow at the source strictly satisfies the burst conditions is also taken into account in our analysis model. The behavior of the packets at the source is shown in detail in Figure 7. 

\begin{figure}[htbp]
\centerline
{
\includegraphics[height=9.4cm,width=7cm]
{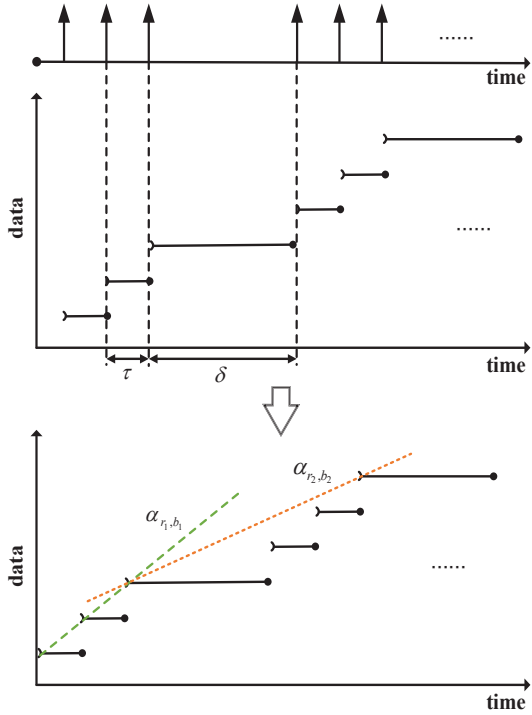}
}
\caption{Arrival process Reproduction of data flow in the real network environment}
\label{fig8}
\end{figure}

A sufficient and necessary condition for generating burst data flow is that the forwarding packet device has a extremely high port rate, which is difficult to achieve in the real network environment. In our model, the periodic burst flow is fed by the server as system input. Ideally, packets should be sent at the same time during each period. However, as shown in the Figure 7, there is a time interval $\tau$ between each two packet sending times during the period, so our previous analysis is not sufficient to meet the existing test conditions. And we define the interval time between the last packet sent in the previous period and the first packet sent in the next period as $\delta$. Thus, we derive that
$$
\frac{n\times l}{(n-1)\tau+\delta}=r\times1Gbps\ ,
\ \tau=\frac{l}{r_p}
$$
where $r$ represents the load of the server, $n$ represents packet burst (that is, $b=nl$) and $r_p$ means the maximum transmission for the fowarding packet device. 

In this way, the new arrival curve needs to be applied to our calculations. The packets of the burst periodic flow with period $T_{p}=(n-1)\tau+\delta$ is sent at times $nT_p$, $nT_p+\tau$ and $nT_p+2\tau$, $n\in \mathbb{N}$. Then it is derived that the cumulative arrival process is $A(t)=\kappa_{l,T_p}(t)$ + $\kappa_{l,T_p}(t-\tau)$ + $\kappa_{l,T_p}(t-2\tau)$. As illustrated in the bottom part of the Figure 8, the best arrival process can be constrained by two linear functions $\alpha_{r_1,b_1}$ and $\alpha_{r_2,b_2}$. Thus, the arrival curve of arrival process $A$ can be derived by
$$
\alpha_{r_1,b_1,r_2,b_2}=A\oslash A=\alpha_{r_1,b_1}\land\alpha_{r_2,b_2},
$$
where
$$
\ r_1=\frac{l}{\tau}, b_1=l, 
$$
$$
r_2=\frac{nl}{T_p}, b_2=nl-r_2\tau(n-1)
$$

Until now, the delay can be computed by the arrival curve $\alpha_{r_1,b_1,r_2,b_2}$ and service curve $\beta_{R,e}$ ($r_2\le R\le r_1$)as
\begin{equation}
\begin{aligned}
d(A,D)&\leqslant h(\alpha_{r_1,b_1,r_2,b_2},\beta_{R,e})\\&=(\frac{l}{R}-\tau)n+e+\tau
\end{aligned}
\end{equation}

By observing the gap between the measurement data and the theoretical calculation value and explaining the reasons for the error, three corrective schemes that can make the theoretical results closer to the measured data values have been proposed in this section. These modifications give new mathematical descriptions of the arrival process with emphasis on universality, strictness and practicality.

\subsection{Comparison Of three modifications}

So far, we have completed the modification of the single node model considering various factors influencing the results of delay analysis in a real network environment. Then on the base of these, we will simulate the time delay on the basis of equation (4), (5) and (6) for comparison. For clarity, the model based on pervasive arrival curve is defined as Model A, the model based on arrival curve of
strict periodic flow is defined as Model B and the model based on strict arrival curve for data flow in actual network environment is defined as Model C. 

As shown in Figure 8, calculations curve based on Model A change with the increase of load and conversely curve based on Model B and Model C is not affected by the load. Theoretically, all measurement data is located below the curve calculated by the model based on Model B with arrival curve of strict periodic flow, however, testing data are scattered on both sides of the black line as in Figure 8. We make an analysis of the causes resulting in this phenomenon and give a conjecture that whether the forwarding package instrument in test achieves strict periodicity of output data package is the hinge on which a periodic flow grows into a pervasive arrival curve $\alpha_{p,l,r,b}$ or more practicable$\alpha_{r_1,b_1,r_2,b_2}$.

As shown in the left side of Figure 8, the testing delay 13.0us has exceeded the calculated delay 9.6us in method on basis of strict periodic flow when the load is 80\%. The red dotted line, which takes into account the real performance of the server and the speed of the port link with $\alpha_{p,l,r,b}$, lie above all discrete testing data points as the final correction result. For the periodic flow, we depict the arrival process by more precise curve $\alpha_{T,b}$ for comparison with the curve based on Model A and Model C. Through arrival curve $\alpha_{p,l,r,b}$ offering some parameter calibration methods which in specific to the rate and burst constrain of the incoming traffic flow as shown in Figure 5, calculating latency changes with load and all measurement data satisfy the delay boundary based on equation (4). 

From the observation result in Figure 8 and Figure 9, we found that the delay calculated by model C is closer to the measured data when the packet length is larger, however, the measured data can be better constrained by the delay calculated by model A. As the number of bursts increases, the calculation delay of model C does not change much as illustrate in Figure 9. Thus, the pervasive arrival curve $\alpha_{p,l,r,b}$ is generally more broadly applicable across various flow in network environment through experimental data analysis.

\begin{figure*}
\centerline
{
\includegraphics[height=4.8cm,width=15.8cm]
{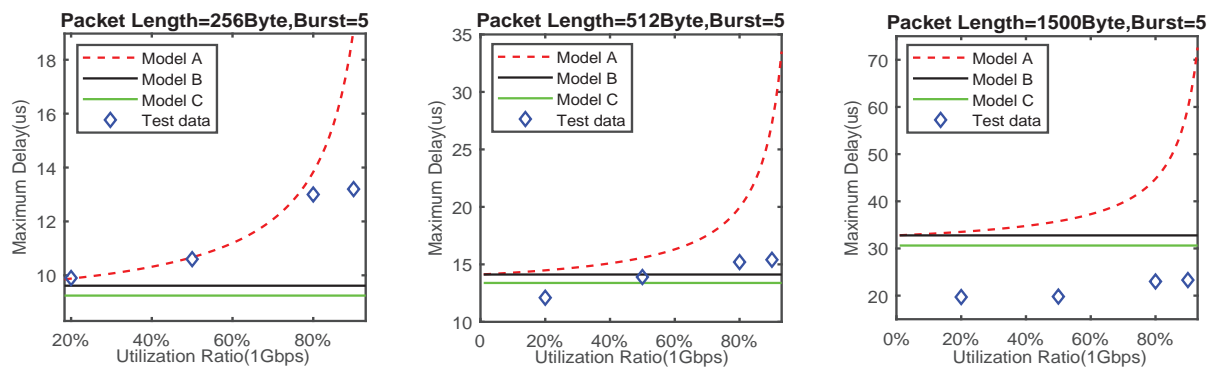}
}
\caption{The guarantee of maximum delay under different data flow intensity   (packet length =256, 512 ,1500Byte, burst = 5 packet length)}
\label{fig8}
\end{figure*}

\begin{figure*}
\centerline
{
\includegraphics[height=4.65cm,width=15.9cm]
{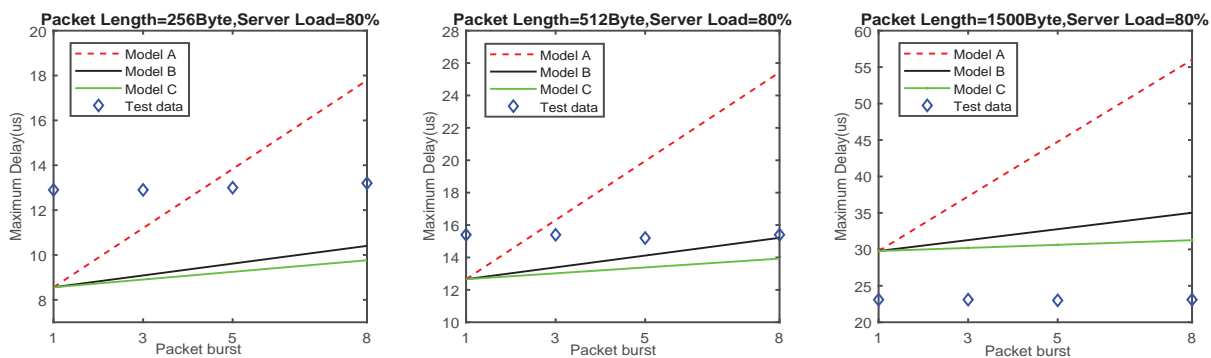}
}
\caption{Impact of Packet burst on Delay (packet length =256, 512 ,1500Byte, server load =80\%)}
\label{fig9}
\end{figure*}

\section{Conclusion}

Although the delay boundary can be derived by classical theory directly, there is still a certain gap between calculated and measured. The classical theory must be further revised according to the actual situation. In this paper, we propose some modification schemes to reduce the deviation between the measured value and the theoretical value. In order to realize the detection of real service capability boundaries of servers, we propose a detection algorithm with the analysis of testing data. Considering the complex internal structure of network entities where each packet is not as expected in fowarding rules, some arrival model is presentd to realize the accurate description of network behavior. The experimental results are obtained through the test environment, and the corresponding theoretical results are obtained based on our analysis. The experimental results are compared with the theoretical results, analyzed and improved.

\end{document}